\documentclass[aps,pra,twocolumn,groupedaddress]{revtex4-1}

\usepackage{graphicx}
\usepackage{amsmath}
\usepackage{graphicx}
\usepackage[colorinlistoftodos]{todonotes}
\usepackage[colorlinks=true, allcolors=blue]{hyperref}
\usepackage{textcomp}
\usepackage{soul}
\usepackage{xcolor}

\begin{document}

\setlength{\abovedisplayskip}{2pt}
\setlength{\belowdisplayskip}{2pt}



\title{Universal Symmetry Breaking Dynamics for the Kerr Interaction of Counter-Propagating Light in Dielectric Ring Resonators}
\author{Michael T. M. Woodley,$^{1,2}$ Jonathan M. Silver,$^{1}$ Lewis Hill,$^{3}$ Fran\c{c}ois Copie,$^{1}$ Leonardo Del Bino,$^{1,2}$ Shuangyou Zhang,$^{1}$ Gian-Luca Oppo,$^{3}$ and Pascal Del'Haye.$^{1}$}

\affiliation{$^{1}$National Physical Laboratory, Hampton Road, Teddington, TW11 0LW, UK,}

\affiliation{$^{2}$Heriot-Watt University, Edinburgh, EH14 4AS, UK,}

\affiliation{$^{3}$Department of Physics, University of Strathclyde, Glasgow, G4 0NG, UK.}





\begin{abstract}
Spontaneous symmetry breaking is an important concept in many areas of physics. A fundamentally simple symmetry breaking mechanism in electrodynamics occurs between counter-propagating electromagnetic waves in ring resonators, mediated by the Kerr nonlinearity. The interaction of counter-propagating light in bi-directionally pumped microresonators finds application in the realisation of optical non-reciprocity (for optical diodes), studies of $\mathcal{PT}$-symmetric systems, and the generation of counter-propagating solitons. Here, we present comprehensive analytical and dynamical models for the nonlinear Kerr-interaction of counter-propagating light in a dielectric ring resonator. In particular, we study discontinuous behaviour in the onset of spontaneous symmetry breaking, indicating divergent sensitivity to small external perturbations. These results can be applied to realise, for example, highly sensitive near-field or rotation sensors. We then generalise to a time-dependent model, which predicts new types of dynamical behaviour, including oscillatory regimes that could enable Kerr-nonlinearity-driven all-optical oscillators. The physics of our model can be applied to other systems featuring Kerr-type interaction between two distinct modes, such as for light of opposite circular polarisation in nonlinear resonators, which are commonly described by coupled Lugiato-Lefever equations.

\end{abstract}


\maketitle


\section{Introduction}
\label{sect_1}
Spontaneous symmetry breaking plays a critical role in the description of many phenomena in physics. In the case of continuous symmetries, it allows for the modelling of magnetism and superconductivity \cite{sc}, as well as the generation of mass via the Higgs mechanism \cite{Higgs}. It also plays a prominent role in systems that exhibit discrete symmetries, which are frequently found in optics, such as time-reversal \cite{T-rev,Xu2014} and parity-time symmetries \cite{PT}, as well as the interplay between two types of symmetry breaking \cite{interplay}. A novel type of discrete symmetry breaking has recently been demonstrated in bi-directionally-pumped whispering-gallery microresonators \cite{ours,Xiao}. In this case, the symmetry violation is caused by an instability whereby, above a threshold pump power, a difference between the intracavity powers in the two counter-propagating directions leads to a splitting between their two resonant frequencies via the Kerr nonlinearity. As a result, small differences between the intracavity circulating powers are amplified. Consequently, the (parity) symmetry of the circulating optical power in the resonator spontaneously breaks. This is due to the fact that the cross-phase-modulation-induced Kerr shift between counter-propagating light waves is different to the self-phase-modulation-induced shift in unidirectional light. Interestingly, this imposes fundamental limits on the attainable power of a standing wave in a dielectric ring resonator. Fig.~1 shows a simple experimental platform for observing this symmetry breaking in a ring resonator.

The theoretical treatment of the Kerr interaction of counter-propagating light was pioneered by Kaplan, Meystre, and collaborators in the early 1980s, in the context of nonlinear effects in Sagnac interferometers \cite{KM81,KM82,WMFK85}. A more complete theoretical basis for light-with-light interaction in ring resonators, and especially of symmetry breaking, is critically needed for a precise understanding of recent work on microresonator-based non-reciprocal devices such as isolators and circulators \cite{iso}, as well as for the dynamics of counter-propagating solitons \cite{QFYang17,ranging1,ranging2}. Here, we present a generalised model that not only captures this symmetry breaking (see Section II), but extends to a universal sensitivity analysis, which will be crucial for future sensing devices based on this effect, such as enhanced rotation sensors and near-field detectors (see Section III). In particular, we determine minimal requirements for the input power to achieve symmetry breaking and provide optimal conditions of operation to counteract the effect of imbalanced pumping conditions in the counter-propagating optical beams. In Section IV, we generalise the model to the time domain, find analytical stability conditions and determine the onset and the frequency of nonlinear oscillations, thus facilitating the creation of an all-optical oscillator. 
 

\begin{figure}[h!]
\centering
\includegraphics[width=0.9\columnwidth]{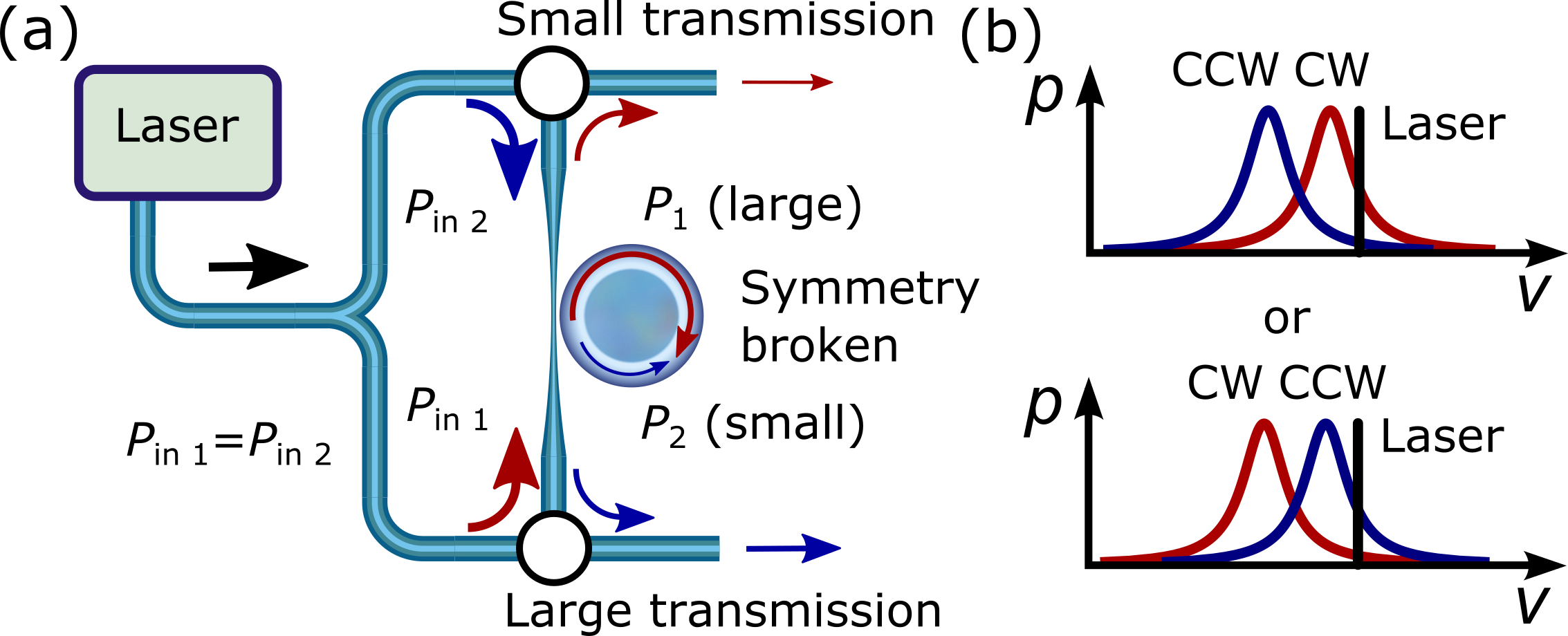}
\caption{A simplified schematic of a ring resonator setup in a symmetry-broken state. Amplified laser light ($\lambda=1550$ nm) is split into two paths of equal powers, which are then sent in opposite directions into a microresonator. (b) Above a critical power threshold, the parity symmetry is broken, characterised by different resonant frequencies in the two directions, such that one direction of light is preferentially coupled into the cavity. Panel (a) shows the case for preferential coupling in the clockwise direction. CW = clockwise, CCW = counterclockwise.}
\end{figure}   

\section{Onset of symmetry breaking}
\label{sect_2}
The basis of our theory of counter-propagating light in a ring resonator is the following dimensionless model, featuring two coupled Lorentzian curves, which results from considering how the Kerr effect modifies the resonant frequency of a cavity in a non-reciprocal fashion \cite{ours}:

\vspace{-3mm}

\begin{equation}\label{eq:dedim}
p_{1,2}=\frac{\tilde{p}_{1,2}}{1+(p_{1,2}+2p_{2,1}-\Delta_{1,2})^{2}}.
\end{equation}
$\tilde{p}_{1,2}$ are the (dimensionless) powers of the pump laser, and $p_{1,2}$ are the powers coupled into the resonator. The apportionment of power coupled into the resonator is determined by the detuning parameters, $\Delta_{1,2}$, which are normalised to the half-linewidth of the resonance. The subscripts 1 and 2 denote clockwise and counterclockwise directions. See Appendix A for more information. Equations \eqref{eq:dedim} are the steady-state, homogeneous solutions to a pair of coupled Lugiato-Lefever equations (see Ref.~\cite{bif} for the uncoupled version). Consequently, this model extends to other nonlinear systems in which Kerr coupling occurs between two distinct modes, such as for opposite circular polarisation states \cite{Haelt_disp, Geddes94}, thus increasing its universality of application. We now need to consider the threshold condition beyond which Eq.~\eqref{eq:dedim} can describe a symmetry-broken regime. 

We demonstrate here that symmetry breaking occurs only above a certain threshold pump power. We consider symmetric pumping conditions by setting $\Delta_{1}=\Delta_{2}=\Delta$ and $\tilde{p}_{1}=\overset{\sim}{p}_{2}=\tilde{p}$, and then examine the number of crossing points between the two Lorentzians, Eq.~\eqref{eq:dedim}, in terms of $p_{1,2}$, to see where this number changes between one and three (two stable, one unstable - indicative of a bistable regime, and hence symmetry breaking). Combining the two Lorentzians gives the following cubic equation:

\begin{equation}\label{eq:cubic}
[p_{1}-p_{2}]\left[(p_{1}^{2}+p_{2}^{2}+p_{1}p_{2})-2\Delta(p_{1}+p_{2})+\Delta^{2}+1\right]=0.
\end{equation}\\
The term in the first set of square brackets is the symmetric solution, whilst the rest is the symmetry-broken solution. The abrupt onset of symmetry breaking is hence described as the discontinuous intersection of a straight line with an ellipse, as shown in Fig.~2. At the points where symmetry breaking occurs, i.e., the points of intersection, $p_{1}=p_{2}=p_{\pm}$, the quadratic part of Eq.~\eqref{eq:cubic} yields

\vspace{-3mm} 

\begin{equation}\label{eq:psplit}
p_{\pm}=\frac{1}{3}\left(2\Delta\pm\sqrt{\Delta^{2}-3}\right),
\end{equation}
for $\Delta\geq\sqrt{3}$. The $p_{-}$ solution gives the coupled power at which the symmetry-broken region opens, and $p_{+}$ gives the coupled power at which it closes. Similar threshold powers apply to polarisation symmetry breaking in ring cavities, as shown in \cite{Haelt_disp}. Coincidentally, Eq.~\eqref{eq:psplit} describes the limits of the region of bistability for the homogeneous, steady-state solution to a Lugiato-Lefever equation for a unidirectional beam of light in a cavity \cite{bif}. The threshold power (per direction) for the same bistability in the case of a standing wave is, in fact, 3 times lower than in the unidirectional case, owing to the extra contribution from cross-phase modulation \cite{KM82}.

\begin{figure}[h!]
\centering
\includegraphics[width=0.8\columnwidth]{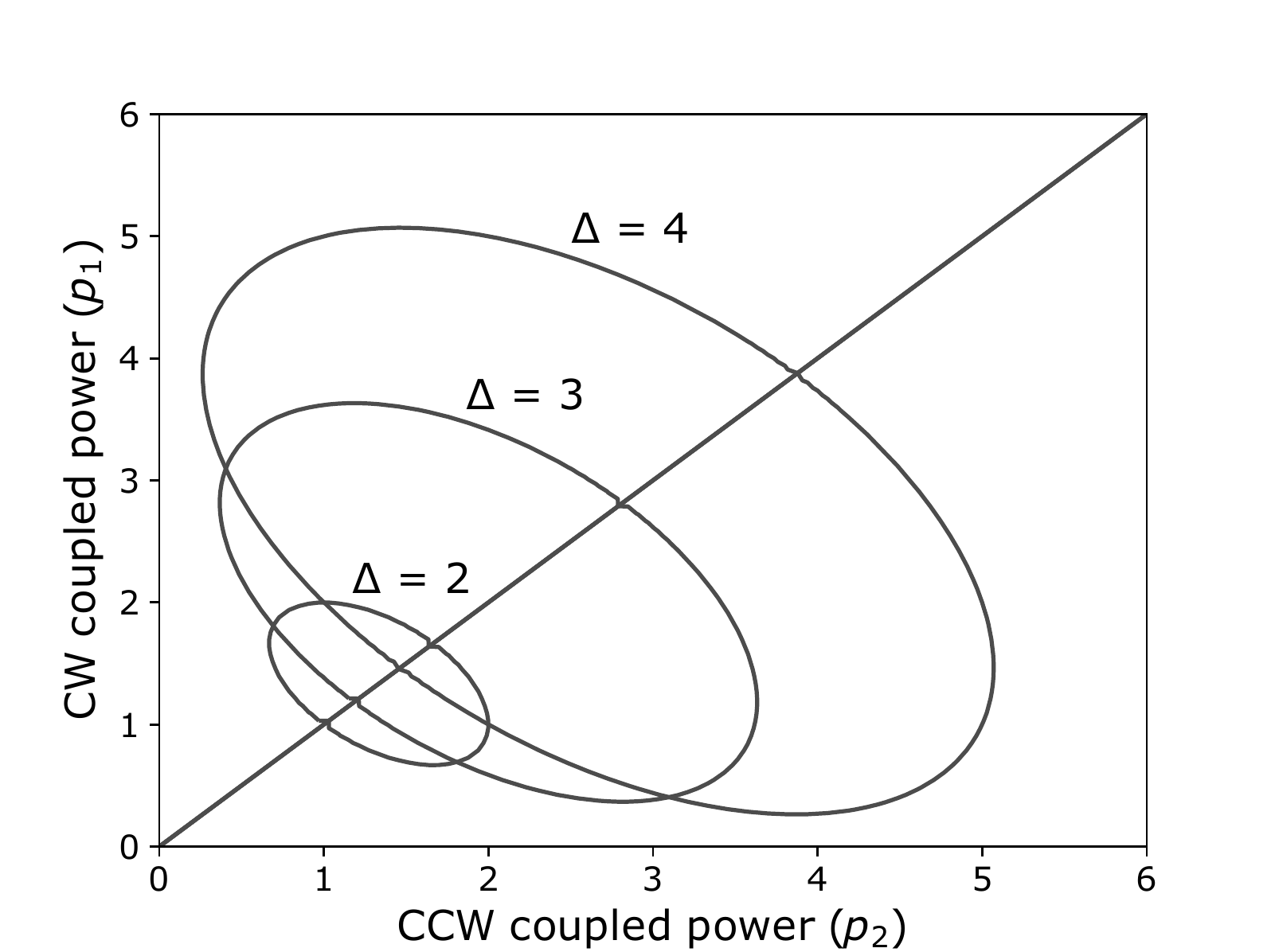}
\caption{Illustration of Eq.~\eqref{eq:cubic}, each ellipse for a different (constant) value of the detuning parameter $\Delta$. Each curve comprises all the values of $p_{1,2}$ at which the two Lorentzians, Eq.~\eqref{eq:dedim} intersect. When $p_{1}\neq p_{2}$, the symmetry is broken, the threshold of which occurs where an ellipse intersects the straight line. These curves are traced out by continuously increasing the pump power, $\tilde{p}$.}
\end{figure}    

In order to find the minimum required pump power for symmetry breaking as a function of detuning only, we first apply Eq.~\eqref{eq:psplit} to Eq.~\eqref{eq:dedim} for $p_{1}=p_{2}=p_{\pm}$ and $\tilde{p}_{1}=\tilde{p}_{2}=\tilde{p}$, to recover the equation found in Ref.~\cite{KM82}:

\begin{equation}\label{eq:thresh}
\tilde{p}_{\pm}=\frac{2}{3}\left[\Delta(3\Delta^{2}-5)\pm(3\Delta^{2}-1)\sqrt{\Delta^{2}-3}\right].
\end{equation}
As for the coupled power, $\tilde{p}_{-}$ gives the pump power at which the symmetry-broken region opens, and $\tilde{p}_{+}$ is where it closes. The threshold pump power, which extends the analysis of Ref.~\cite{KM82}, is then found to be

\vspace{-2mm}

\begin{equation}
\tilde{p}_{\mathrm{thresh.}}=\tilde{p}_{-}\left(\Delta=\frac{5}{\sqrt{3}}\right)=\frac{8}{3\sqrt{3}}\approx 1.54,
\end{equation}
associated with the coupled power of $p=2/\sqrt{3}\approx 1.15$.

\section{Nonlinear enhancement of sensitivity}
\label{srct_3}
We have so far considered balanced pump powers and equal detunings in both directions of propagation, in order to investigate symmetry breaking. We now consider the most general case of Eq.~\eqref{eq:dedim}, for $\tilde{p}_{1}\neq\tilde{p}_{2}$ and $\Delta_{1}\neq\Delta_{2}$. The latter may be caused by using the resonator as a sensor, whereby a change in the local environment around the resonator perturbs its optical modes \cite{sens1,sens2}, for example by inducing a small resonant frequency splitting. This can come from rotating the resonator (the Sagnac effect) or from interacting with its evanescent field. This splitting can then be magnified by the Kerr nonlinearity, allowing for enhancement of the sensitivity, theoretically down to the shot noise limit. This idea has been explored in the case of symmetric pumping in Refs.~\cite{KM81, WangSearch2014, WangSearch2015}.

\textit{Divergent sensitivity.}---Sensitivity, in this context, is defined as the rate of change of the coupled power with respect to the detuning. These partial derivatives are calculated to be

\begin{equation}\label{eq:x2}
\frac{\partial p_{1,2}}{\partial\Delta_{1,2}}=\frac{(1+X_{2,1})}{(1+X_{1})(1+X_{2})-4},
\end{equation}

\begin{equation}\label{eq:y2}
\frac{\partial p_{1}}{\partial\Delta_{2}}=\frac{\partial p_{2}}{\partial\Delta_{1}}=-\frac{2}{(1+X_{1})(1+X_{2})-4},
\end{equation}
wherein

\begin{equation}\label{eq:tune}
X_{1,2}=\frac{1+(p_{1,2}+2p_{2,1}-\Delta_{1,2})^{2}}{2p_{1,2}(p_{1,2}+2p_{2,1}-\Delta_{1,2})}.
\end{equation}
See Appendix B for details of the derivation of these expressions. By inspection, the sensitivity diverges for

\begin{equation}\label{eq:sens}
(1+X_{1})(1+X_{2})=4.
\end{equation}
Eq.~\eqref{eq:sens} is the universal condition for maximally-enhanced sensitivity to a small perturbation to the resonator; it defines a closed boundary within the symmetry-broken regime, and is illustrated in Fig.~3. Unlike previous sensitivity analyses, such as in Refs.~\cite{KM81, WangSearch2014, WangSearch2015}, Eq.~\eqref{eq:sens} is valid for arbitrary pumping and detuning conditions, and is therefore much more general. As will be demonstrated later, Eq.~\eqref{eq:sens} defines the onset of instability in the system. For given pump and circulating powers, this condition defines the detunings for which sensitivity diverges. However, since the region enclosed by this boundary is unstable (see the stability analysis that follows), we find that only one particular point on this boundary is useful for nonlinear enhancement of a sensor, referred to as the critical point.

\begin{figure}[h!]
\centering
\includegraphics[width=0.7\columnwidth]{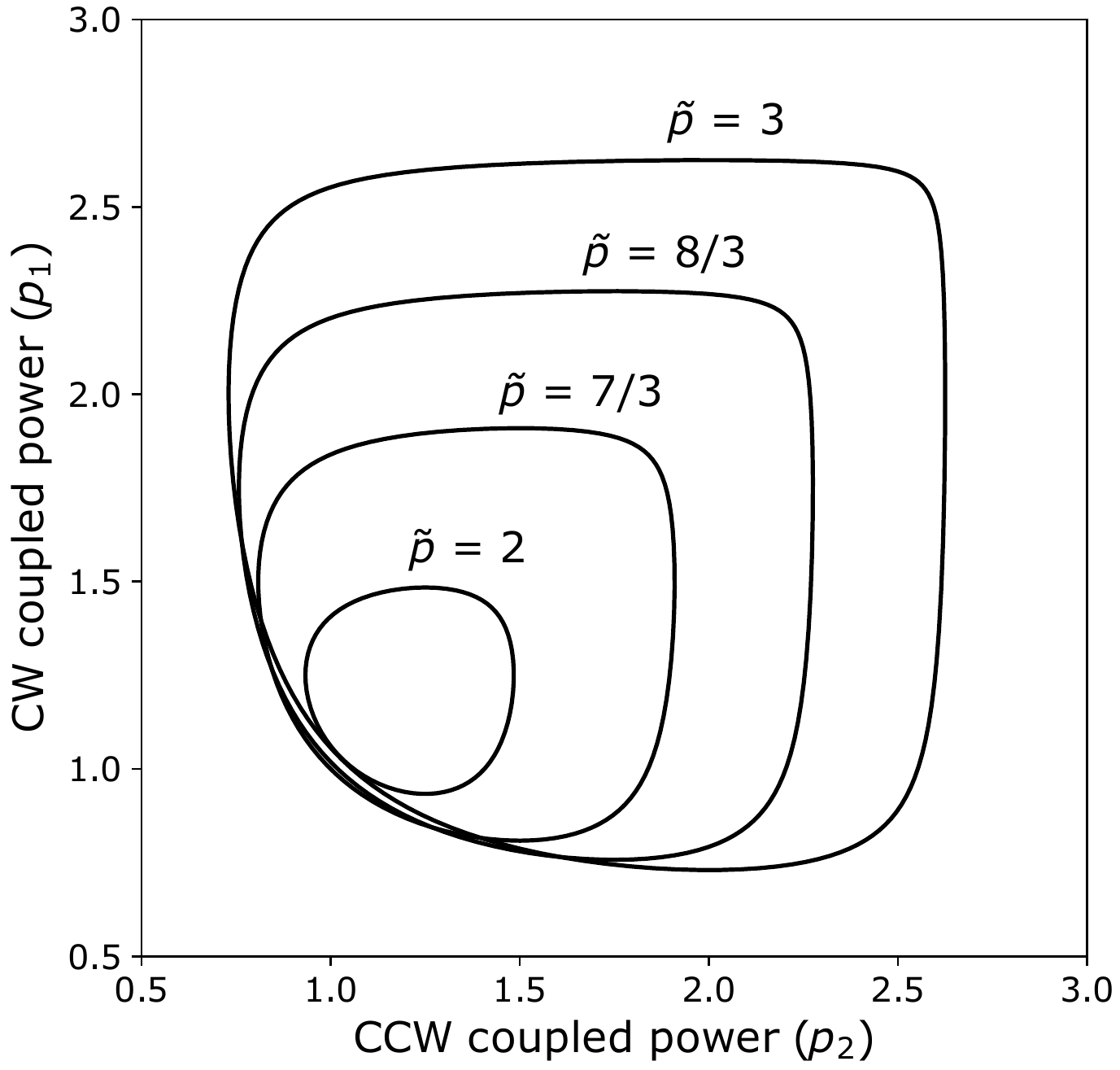}
\caption{Illustration of Eq.~\eqref{eq:sens}, in the case of balanced pump powers, $\tilde{p}_{1}=\tilde{p}_{2}=\tilde{p}$. Each curve corresponds to a different choice of pump powers. Each pair of circulating powers, $(p_{1},p_{2})$, corresponds to a particular pair of detunings, $(\Delta_{1}, \Delta_{2})$.}
\end{figure}

 \textit{Critical point.}---In an experiment, it can be difficult to achieve balanced pump powers. Consequently, it is very useful to know how to achieve divergent sensitivity by compensating imbalanced pump powers with imbalanced detunings. This can be done by satisfying a particular condition that recovers the discontinuous pitchfork-like behaviour observed in symmetry breaking, despite imbalanced pumping conditions. This condition is known as the critical point, and is a more general condition than just the thresholds for symmetry breaking. The critical point sits on the unstable boundary, shown in Fig.~3, and is defined by the condition $X_{1}=X_{2}=1$. Informally, this can be thought of as the condition that causes the sensitivity of the clockwise and counterclockwise coupled powers to diverge at equal and opposite rates. In the symmetric case, i.e., for $\tilde{p}_{1}=\tilde{p}_{2}=\tilde{p}$, $\Delta_{1}=\Delta_{2}=\Delta$, and $p_{1}=p_{2}=p$, this constraint allows us to recover the original condition for symmetry breaking, Eq.~\eqref{eq:psplit}. In this way, the critical point condition, as a special case of Eq.~\eqref{eq:sens}, constitutes a generalisation of the sensitivity analyses shown in Refs.~\cite{KM81, WangSearch2014, WangSearch2015}. Consequently, divergent sensitivity to perturbations can be accessed even in the case of imbalanced pump powers and unequal detunings. We have numerically verified this condition for the critical point. We note that the sensitivity to changes in the pump powers also diverges for this same critical point condition.

\section{Time-dependent model}
\label{sect_4}
We observe that the coupled Lorentzians in Eq.~\eqref{eq:dedim} are, in fact, the steady-state solutions to a pair of time-dependent coupled mode equations. By introducing normalised electric fields $p_{1}=|e_{1}|^{2}$, $p_{2}=|e_{2}|^{2}$, $\tilde{p}_{1}=|\tilde{e}_{1}|^{2}$, $\tilde{p}_{2}=|\tilde{e}_{2}|^{2}$, the resulting equations take the form

\begin{equation}\label{eq:LLE}
\dot{e}_{1,2}=\tilde{e}_{1,2}-[1+i(|e_{1,2}|^{2}+2|e_{2,1}|^{2}-\Delta_{1,2})]e_{1,2},
\end{equation}\\
where the dot signifies the time derivative, and the time has been normalised by the photon lifetime in the cavity. Consequently, the theory shown in the first part of the paper -- symmetry breaking and the sensitivity analysis -- can be subsumed as the steady state properties of a time-dependent theory.

\begin{figure}[h!]
\centering
\includegraphics[width=0.96\columnwidth]{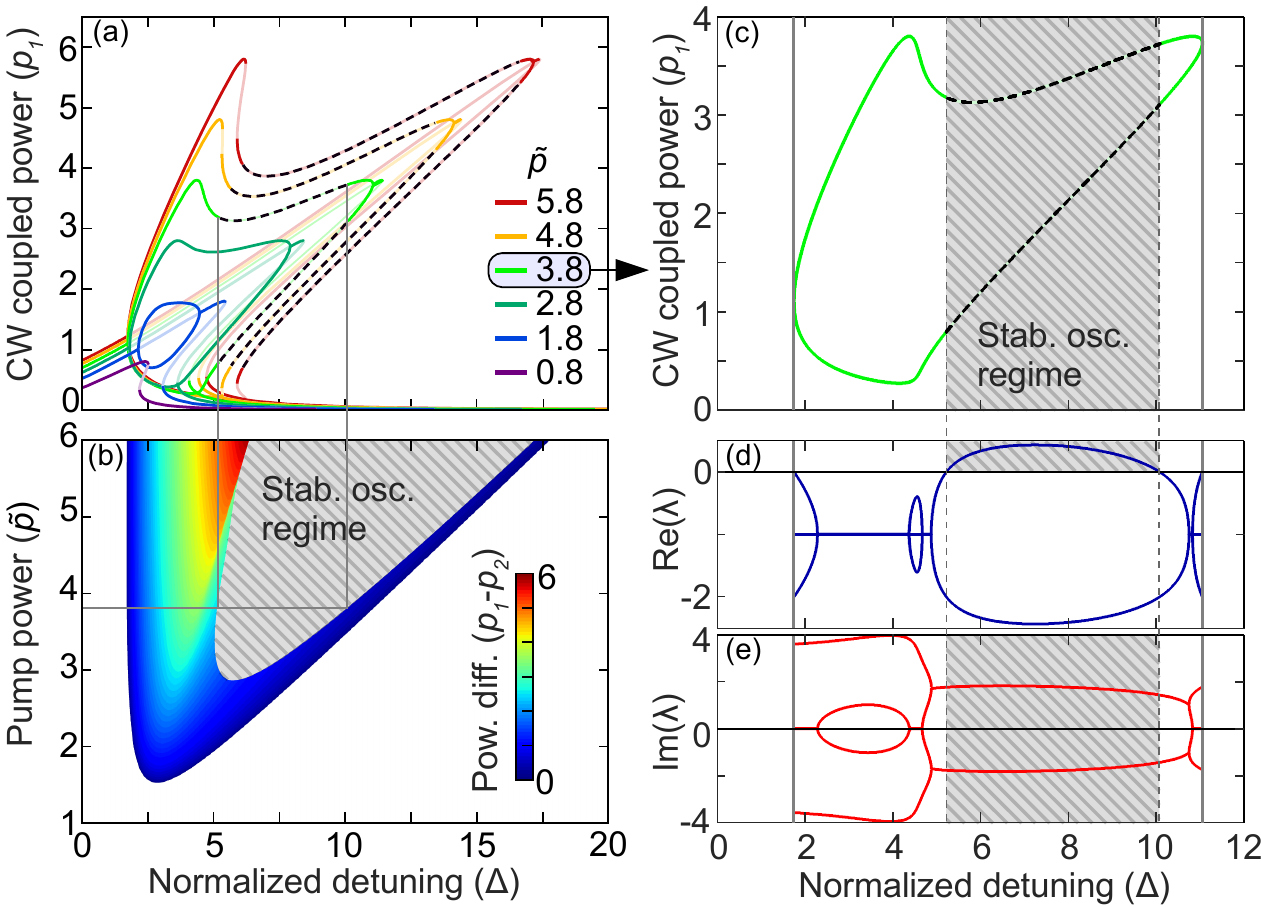}
\caption{Symmetry-broken solutions under balanced pumping. (a) Tilted resonances for various pump powers, showing the symmetry-broken region. Dark solid curves indicate stable solutions, faint curves show unstable solutions, and dashed curves correspond to oscillatory behaviour. (b) Amplitude of the difference in the coupled powers. The grey area corresponds to time-oscillating solutions (white denotes symmetric states). (c) Isolated curve for $\tilde{p}=3.8$, from (a). (d) Real part of the stability eigenvalue, $\lambda$. (e) Imaginary part of the stability eigenvalue. Note that there always exists at least one eigenvalue with non-zero imaginary part, implying strong susceptibility to oscillations.}
\end{figure}

\textit{Stability analysis and oscillations.}--- Small perturbations of the electric field may grow or shrink with time. We investigate this by defining $e_{1}=e_{\mathrm{s}1}+\epsilon_{1}$ and $e_{2}=e_{\mathrm{s}2}+\epsilon_{2}$, where $\epsilon_{1,2}$ are small perturbations on the steady-state solutions, $e_{\mathrm{s}1,\mathrm{s}2}$. For simplicity, but without loss of generality, we adjust the phases of $\tilde{e}_{1,2}$ such that $e_{\mathrm{s}1,\mathrm{s}2}$ are real. To characterise the time evolution of the resulting system of four equations, we form a $4\times 4$ matrix, which has the following eigenvalues:   

\begin{equation}\label{eq:matrixev}
\lambda=-1\pm\frac{\sqrt{-A_{1}B_{1}-A_{2}B_{2}\pm S}}{\sqrt{2}},
\end{equation}

\begin{equation}\label{eq:S}
S=\sqrt{(A_{1}B_{1}-A_{2}B_{2})^{2}\!+4A_{1}A_{2}C^{2}},
\end{equation}
in which $A_{1}=e_{\mathrm{s}1}^{2}+2e_{\mathrm{s}2}^{2}-\Delta_{1}$, $A_{2}=2e_{\mathrm{s}1}^{2}+e_{\mathrm{s}2}^{2}-\Delta_{2}$, $B_{1}=3e_{\mathrm{s}1}^{2}+2e_{\mathrm{s}2}^{2}-\Delta_{1}$,
$B_{2}=2e_{\mathrm{s}1}^{2}+3e_{\mathrm{s}2}^{2}-\Delta_{2}$, and $C=4e_{\mathrm{s}1}e_{\mathrm{s}2}$. The $\pm$ signs are independent, giving four distinct eigenvalues. See Appendix C for more details. Qualitative changes in the eigenvalue describe transitions to different kinds of time-dependent behaviour: if $\lambda$ is a positive real number, the solution is a real, growing exponential, and the perturbed system will become unstable. Such unstable solutions occur for
\vspace{2mm}
\begin{equation}\label{eq:unstable}
(1+X_{1})(1+X_{2})<4,
\end{equation}
i.e., within the region enclosed by each of the curves in Fig.~3.

From Eq.~\eqref{eq:matrixev} and the conditions of their existence, we have verified that it is not possible to have four real eigenvalues for the linear stability of the symmetry-broken solutions. This means that these solutions are strongly susceptible to either damped or sustained oscillations, since at least one stability eigenvalue is complex for experimentally relevant values of the parameters. Intuitively, oscillations are expected because of the simultaneous presence of two symmetry-broken solutions, under exchange of the indices $(1,2)$ in Eq.~\eqref{eq:LLE}. In Fig.~4 (e) we show, for example, that for the chosen value of the pump $\tilde{p}=3.8$, there is always at least one eigenvalue with non-zero imaginary part. In the case of negative real parts of the stability eigenvalues (between the vertical solid and dashed lines in Fig.~4 (c)--(e)) one observes damped oscillations where nonlinear resonances can be excited by suitable modulations of the pumps. When $S$ in Eq.~\eqref{eq:S} is purely imaginary, we have four complex eigenvalues where the angular frequency, $\Omega$, and the growth rate, $R$, are given respectively by

\begin{equation}\label{eq:angfreq}
\Omega=\pm\sqrt{\frac{1}{2}\sqrt{A_{1}A_{2}(B_{1}B_{2}-C^{2})}+\frac{1}{4}(A_{1}B_{1}+A_{2}B_{2})},
\end{equation}

\begin{equation}\label{eq:growth}
R=-1\pm\sqrt{\frac{1}{2}\sqrt{A_{1}A_{2}(B_{1}B_{2}-C^{2})}-\frac{1}{4}(A_{1}B_{1}+A_{2}B_{2})} \; .
\end{equation}
Here, again, the independent $\pm$ signs give rise to four different complex solutions. Oscillations due to the Kerr nonlinearity have been previously noted to occur in systems of microcavities, including those with multiple coupled resonators \cite{Maes2009, Abdollahi2014, Dumeige2015}, and in a single resonator, in which linear coupling is achieved between the two cavity modes via Rayleigh backscattering \cite{Dumeige2011}. In contrast, our analysis, which uses direct bi-directional pumping of a single ring resonator, yields several new insights about these nonlinear oscillations in the presence of counter-propagation. For example, exact analytical expressions for the angular frequencies and growth rates of the oscillations, Eqs.~\eqref{eq:angfreq} and \eqref{eq:growth}, are available, and our linear stability analysis is valid for arbitrary choices of pump powers and detunings. We have also demonstrated that counter-propagating light is strongly susceptible to oscillations since there can never be four real eigenvalues of the linear stability in the symmetry-broken regime. Finally, stable and chaotic oscillations are here associated with a collision of two Hopf bifurcations, as demonstrated below.

When $R$ becomes positive, two Hopf bifurcations (forward and backward, when changing $\Delta$) of the symmetry-broken solutions occur, allowing for non-decaying stable oscillations. We illustrate these oscillations as the dashed curves and shaded grey regions in Fig.~4. In frame (e) of this figure, we plot the corresponding frequencies of oscillation, $\Omega$, from Eq.~\eqref{eq:angfreq} when changing the detuning. We anticipate that this bifurcation will permit the construction of a microresonator-based all-optical oscillator, featuring periodic energy exchange between the two directions, in the nanosecond regime.  
\begin{figure}[h!]
\centering
\includegraphics[width=0.96\columnwidth]{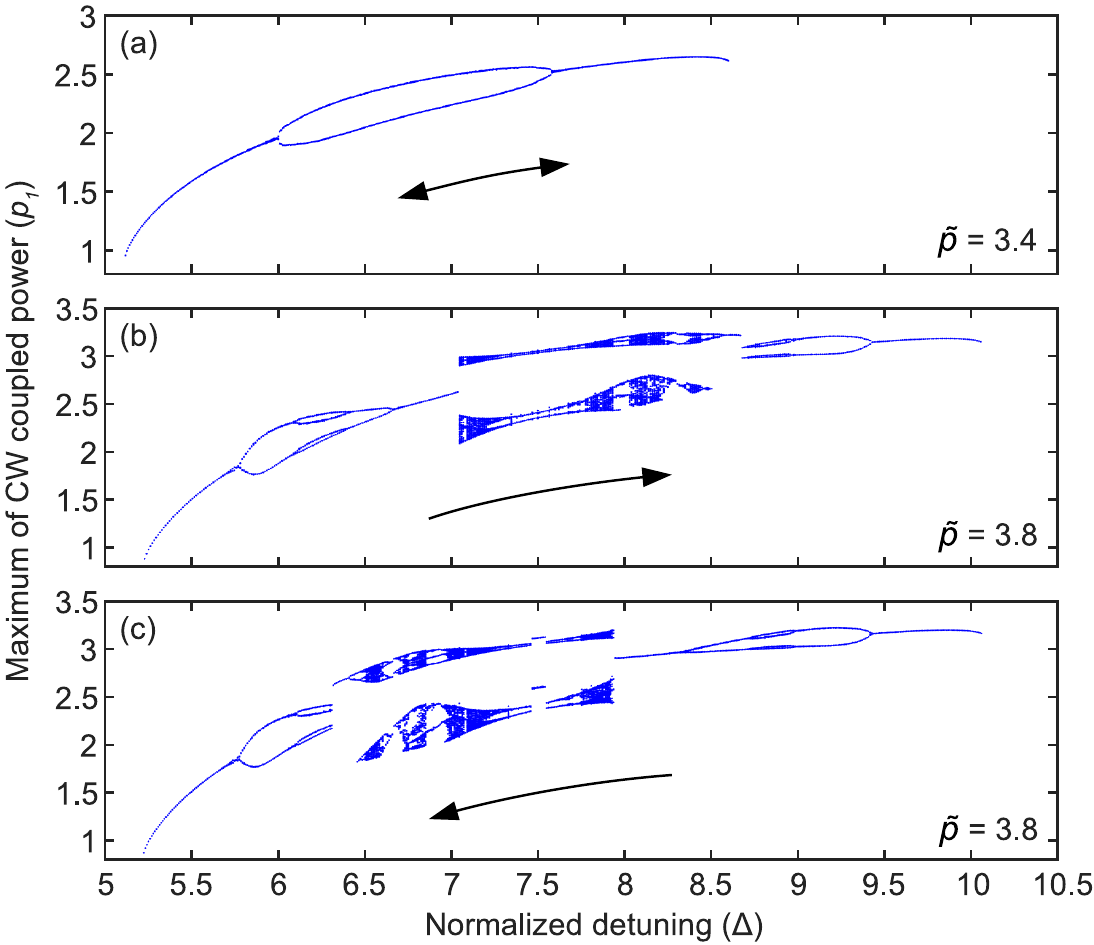}
\caption{Detuning scans of the oscillatory regimes of Eq.~\eqref{eq:LLE} between two Hopf bifurcations (lower solutions of Fig.~4 (a)). The field intensity is sampled at its maximum during the oscillation. A single value of $p_{1}$ for a given $\Delta$ means that only one (maximal) value of $p_{1}$ intersects the Poincar\'{e} section during its trajectory with a single periodicity. In contrast, two values of $p_{1}$ for a specified $\Delta$ means that the maximum of $p_{1}$ is cycling between two distinct values with an overall period that has doubled. Further period-doublings are observed for larger values of $\Delta$, which eventually transition into chaos. The arrows indicate the direction of scanning the detuning. (a)  Scan for $\tilde{p}=3.4$ (showing no dependence on direction). Forward (b) and backward (c) scans for $\tilde{p}=3.8$.}
\end{figure}

We have verified these predictions by direct numerical integration of Eq.~\eqref{eq:LLE}. For each configuration specified by $\Delta$ and $\tilde{p}$ in the oscillatory regime, we sample the trajectories of $p_{i}$, where $i=1,2$, by evaluating the Poincar\'{e} section corresponding to local maxima of $p_{i}$ where the first derivative is zero and the second derivative is negative. In this way, we can monitor the changes in the number of maxima of $p_{i}$ over successive cycles as displayed in Fig.~5. A maximum value of $p_{1}$ is specified by the following condition:

\begin{equation}
p_i=|e_i|^2=\tilde{e}_i \, {\rm Re}(e_i),
\end{equation}
where ${\rm Re}(e_i)$ is the real part of the complex field $e_i$, and we have now adjusted the phases such that $\tilde{e_{i}}$ is real. This method allows us to generate dynamical scans of the oscillatory regime whilst varying $\Delta$. These scans are presented in Fig.~5 for $\tilde{p}=3.4$, $\tilde{p}=3.8$ while changing $\Delta$ from 5 to 11, and for $\tilde{p}=3.8$ while changing $\Delta$ in reverse from 11 to 5. Each point in Fig.~5 corresponds to a nonlinear oscillation of the symmetry-broken output. The numerically-derived oscillation frequencies are in remarkably good agreement with those predicted by Eq.~\eqref{eq:angfreq}.

Forward and backward Hopf bifurcations are clearly visible at the beginning and the end of the scans. For values of input pumps just above the critical value of $\tilde{p}=2.87$ where the Hopf bifurcations appear, regular oscillations occur at frequencies around four times the decay rate of the ring resonator, i.e. the inverse of the photon lifetime. We note that the point $(\tilde{p},\Delta)=(2.87,5.8)$ corresponds to a co-dimension 2 bifurcation where the forward and backward Hopf bifurcations collide. Normal forms and unfoldings of double Hopf bifurcations provide the framework to establish the full dynamical behaviour of the symmetry-broken solutions \cite{knobloch88}. In Fig.~5 (a) we show that forward and backward period doubling bifurcations occur close to the co-dimension 2 point ($\tilde{p}=3.4$). By increasing the pump parameter, further period doubling bifurcations, deterministic chaos, collision of Feigenbaum cascades \cite{oppo84}, and crises emerge, as shown in Fig.~5 (b) and (c) for $\tilde{p}=3.8$. We also observe bistable behaviour between these dynamical regimes, as shown in Fig.~5 (b) and (c), where the direction of the detuning scan is reversed.

\section{Conclusions}
\label{sect_5}
We have presented analytical and dynamical models for the interaction between counter-propagating light in a dielectric ring resonator. The mixture of self- and cross-phase modulation from the Kerr effect results in dramatic changes in behaviour: notably, spontaneous symmetry breaking and the onset of nonlinear oscillations -- the latter holds promise for the development of highly controllable, on-chip, all-optical oscillators. We have also derived the universal condition for divergent sensitivity to perturbations in the cases of both balanced and imbalanced system conditions. This closed boundary of divergent sensitivity with respect to the laser detunings marks the transition between stable and unstable symmetry-broken solutions in a coupled, homogeneous system. The critical point lying on this boundary makes possible a variety of enhanced sensors for detecting rotations or near-field disturbances (e.g. biomolecules). In the dynamical regime, we have obtained analytical expressions of the stability eigenvalues of the symmetry-broken solutions, thus providing thresholds and frequencies of the oscillatory regimes. Collisions of forward and backward Hopf bifurcations when scanning the detuning have been identified as the leading mechanism for the onset of nonlinear oscillations and chaos in counter-propagating ring resonators. In order to cover a wide range of experimental configurations, the scope of our analysis can be extended to systems featuring other forms of Kerr-nonlinearity-mediated symmetry breaking such as the interaction of light states of different polarisations, which would enable the development of optically controllable polarisation filters. In addition, this model is applicable to systems with different coefficients of self- and cross-phase modulation, such as semiconductors and liquid crystals. 

\vspace{3mm}

\begin{acknowledgements}
\noindent We acknowledge financial support from: H2020 Marie Sklodowska-Curie Actions
(MSCA) (748519, CoLiDR); National Physical Laboratory Strategic Research; H2020 European  Research Council (ERC) (756966, CounterLight); Engineering and Physical Sciences Research Council (EPSRC). L. H. acknowledges additional support from the EPSRC DTA Grant No. EP/M506643/1. S. Z. acknowledges funding through the MULTIPLY Horizon 2020 Marie Sklodowska-Curie grant (GA-713694).
\end{acknowledgements}

\appendix

\section{Obtaining the dimensionless model}

\noindent Basic definitions:
\begin{itemize}
\item $\omega_{\mathrm{res}}$ is the cavity resonant frequency;
\item $\gamma_{0}$ is the intrinsic half-linewidth of the cavity resonance;
\item $\kappa$ is the coupling half-linewidth;
\item $\Delta f_{\mathrm{FSR}}$ is the free spectral range of the optical mode family in question;
\item $\delta_{1,2}$ are the differences in frequency (detunings) between the laser and the non-Kerr-shifted clockwise and counterclockwise resonance frequencies of the ring resonator;
\item $A_{\mathrm{eff}}$ is the effective cross-sectional area of the optical mode;
\item $n_{0,2}$ are, respectively, the linear and nonlinear refractive indices of the dielectric resonator. 
\end{itemize}

The action of the Kerr effect on the resonant frequencies of counter-propagating modes in a ring resonator has been described in the SI to Ref.~\cite{ours}, in which the direction-dependent changes in refractive index induced by the Kerr effect (hereafter referred to as the `Kerr-shift') are incorporated into the effective detuning of a Lorentzian intensity profile. Since the Kerr-shift generated by counter-propagating light is different to that produced by a single beam of light, a model of two coupled Lorentzian intensity profiles emerges, which takes the following form:

\begin{equation}\label{eq:dim}
P_{\mathrm{1,2}}=\frac{\eta P_{\mathrm{in}\,1,2}}{1+ \left[ \dfrac{1}{P_{0}}(P_{\mathrm{1,2}}+2P_{\mathrm{2,1}})-\frac{\delta_{\mathrm{1,2}}}{\gamma} \right ]^{2}},
\end{equation}

\noindent in which the $2P_{2,1}$ term is the counter-propagating contribution to the Kerr-shift (a manifestation of cross-phase modulation). The following definitions are presented in the SI to Ref.~\cite{ours}, but are repeated here for convenience: $P_{1}=2\pi P_{\mathrm{circ, CW}}/\mathcal{F}_{0}$ and $P_{2}=2\pi P_{\mathrm{circ, CCW}}/\mathcal{F}_{0}$ are the coupled powers (where $\mathcal{F}_{0}\equiv\pi\Delta f_{\mathrm{FSR}}/\gamma_{0}$ is the intrinsic finesse of the resonator, and $P_{\mathrm{circ,CW}}$ and $P_{\mathrm{circ,CCW}}$ are the clockwise and counterclockwise circulating powers), $P_{\mathrm{in,CW}}$ and $P_{\mathrm{in,CCW}}$ are the clockwise and counterclockwise incident powers, $\eta=4\kappa\gamma_{0}/\gamma^{2}$ is the coupling efficiency ($\gamma=\gamma_{0}+\kappa$ is the loaded half-linewidth), and $P_{0}=\pi n_{0}A_{\mathrm{eff}}/(Q\mathcal{F}_{0}\,n_{2})$ is the characteristic coupled power at which Kerr nonlinear effects occur (wherein $Q=\omega_{\mathrm{res}}/2\gamma$ is the loaded quality factor).  Note that the above definitions for the powers are valid only in the case of critical coupling into the resonator. The subscripts 1 and 2 denote the clockwise and counterclockwise directions of propagation. For simplicity, we now cast this model into dimensionless form by defining the following new quantities: the normalised detunings, $\Delta_{1}=\delta_{\mathrm{1}}/\gamma$, $\Delta_{2}=\delta_{\mathrm{2}}/\gamma$; the clockwise and counterclockwise pump powers, $\tilde{p}_{1}=\eta P_{\mathrm{in, CW}}/P_{0}$, $\tilde{p}_{2}=\eta P_{\mathrm{in, CCW}}/P_{0}$; and the clockwise and counterclockwise coupled powers, $p_{1}=P_{\mathrm{1}}/P_{0}$, $p_{2}=P_{\mathrm{2}}/P_{0}$. In this way, we obtain the dimensionless model, Eq.~\eqref{eq:dedim}.

\section{Sensitivity analysis}
The two tunable parameters in this model are the normalised detunings and the pump powers; changing these will modulate the coupled powers within the cavity. In experiments that exploit symmetry breaking, the clockwise and counterclockwise pump powers are often fixed, and the detunings are varied to scan into and out of the broken symmetry state. In its current form, Eq.~(\ref{eq:dedim}) does not explicitly reflect this, so partial derivatives should be taken with respect to the normalised detunings, to make it clear that the pump powers are indeed constant. Taking partial derivatives of Eq.~(\ref{eq:dedim}) gives

\begin{equation}
\frac{\partial p_{x}}{\partial\Delta_{x}}=\mp\frac{2p_{x}^{2}}{\tilde{p}_{x}}\sqrt{\frac{\tilde{p}_{x}}{p_{x}}-1}(\frac{\partial p_{x}}{\partial\Delta_{x}}+2\frac{\partial p_{y}}{\partial\Delta_{x}}-1),
\end{equation}

\begin{equation}
\frac{\partial p_{y}}{\partial\Delta_{x}}=\mp\frac{2p_{y}^{2}}{\tilde{p}_{y}}\sqrt{\frac{\tilde{p}_{y}}{p_{y}}-1}(\frac{\partial p_{y}}{\partial\Delta_{x}}+2\frac{\partial p_{x}}{\partial\Delta_{x}}),
\end{equation}

\vspace{3mm}

\noindent in which we have also used Eq.~(\ref{eq:dedim}) to simplify the expressions. Rearranging for the rate of change of the (clockwise or counterclockwise) coupled power with respect to the (clockwise or counterclockwise) normalised detuning results in

\begin{equation}
\frac{\partial p_{x}}{\partial\Delta_{x}}=\frac{1-2\dfrac{\partial p_{y}}{\partial \Delta_{x}}}{1\pm\dfrac{\tilde{p}_{x}}{2p_{x}^{3/2}\sqrt{\tilde{p}_{x}-p_{x}}}},
\end{equation}

\begin{equation}
\frac{\partial p_{y}}{\partial\Delta_{x}}=-\frac{2\dfrac{\partial p_{x}}{\partial\Delta_{x}}}{1\pm\dfrac{\tilde{p}_{y}}{2p_{y}^{3/2}\sqrt{\tilde{p}_{y}-p_{y}}}}.
\end{equation}

\noindent Substituting one equation into the other, and writing the expressions in terms of subscripts (1,2), results in Eqs.~\eqref{eq:x2} and \eqref{eq:y2}. It is worth noting that Eq.~\eqref{eq:tune} may also be expressed as

\begin{equation}
X_{1,2}=\pm\frac{\tilde{p}_{1,2}}{2p_{1,2}^{3/2}\sqrt{\tilde{p}_{1,2}-p_{1,2}}}.
\end{equation}

\noindent The critical point condition, $X_{1,2}=1$, hence gives us the following quadratic in $\tilde{p}_{1,2}$:

\begin{equation}
\tilde{p}_{1,2}^{2}-4p_{1,2}^{3}\tilde{p}_{1,2}+4p_{1,2}^{4}=0,
\end{equation}

\noindent which solves as:

\begin{equation}
\tilde{p}_{1,2\,\pm}=2p_{1,2}^{2}\left(p_{1,2}\pm\sqrt{p_{1,2}^{2}-1}\right),
\end{equation}

\noindent with $\tilde{p}_{1,2\,-}$ having a local minimum of $8/(3\sqrt{3})$ at $p_{1,2}=2/\sqrt{3}$, as presented earlier in the article.

By setting $X_{1,2}=1$, and then restoring symmetry by imposing $\tilde{p}_{1}=\tilde{p}_{2}=\tilde{p}$, $\Delta_{1}=\Delta_{2}=\Delta$, and $p_{1}=p_{2}=p$, we obtain

\begin{equation}
3p^{2}-4\Delta p+\Delta^{2}+1=0,
\end{equation}

\noindent which has, as its solution, the threshold coupled powers for spontaneous symmetry breaking given by Eq.~\eqref{eq:cubic}.

\section{Stability analysis}

Defining $e_{1}=e_{\mathrm{s}1}+\epsilon_{1}$ and $e_{2}=e_{\mathrm{s}2}+\epsilon_{2}$, where $\epsilon_{1}$ and $\epsilon_{2}$ are infinitesimal perturbations on the steady-state solutions $e_{\mathrm{s}1}$ and $e_{\mathrm{s}2}$, we obtain the following coupled mode equations for $\epsilon_{1}$ and $\epsilon_{2}$:

\vspace{-3mm}

\begin{eqnarray}
\dot{\epsilon}_{1}=-[1+i(2|e_{\mathrm{s}1}|^{2}+2|e_{\mathrm{s}2}|^{2}-\Delta_{1})]\epsilon_{1} \nonumber \\ -i(e_{\mathrm{s}1}\epsilon_{1}^{*}+2e_{\mathrm{s}2}\epsilon_{2}^{*}+2e_{\mathrm{s2}}^{*}\epsilon_{2})e_{\mathrm{s}1},
\end{eqnarray}

\begin{eqnarray}
\dot{\epsilon}_{2}=-[1+i(2|e_{\mathrm{s}1}|^{2}+2|e_{\mathrm{s}2}|^{2}-\Delta_{2})]\epsilon_{2} \nonumber \\ -i(e_{\mathrm{s}2}\epsilon_{2}^{*}+2e_{\mathrm{s}1}\epsilon_{1}^{*}+2e_{\mathrm{s1}}^{*}\epsilon_{1})e_{\mathrm{s}2}.
\end{eqnarray}

\noindent Here, without loss of generality, we adjust the phases of $\tilde{e}_{1}$ and $\tilde{e}_{2}$ so that $e_{s1}$ and $e_{s2}$ are real. Letting $x_{1,2}$ and $y_{1,2}$ be the real and imaginary parts, respectively, of $\epsilon_{1,2}$, we obtain 

\begin{equation}
\begin{pmatrix}
\dot{x}_{1} \\
\dot{y}_{1} \\
\dot{x}_{2} \\
\dot{y}_{2}
\end{pmatrix}
=
\begin{pmatrix}\label{eq:thematrix}
-1 & A_{1} & 0 & 0 \\
-B_{1} & -1 & -C & 0 \\
0 & 0 & -1 & A_{2} \\
-C & 0 & -B_{2} & -1 
\end{pmatrix}
\begin{pmatrix}
x_{1} \\
y_{1} \\
x_{2} \\
y_{2}
\end{pmatrix}.
\end{equation}

\vspace{3mm}

\noindent $A_{1,2}$, $B_{1,2}$, and $C$ are defined in Section IV, and the eigenvalues of Eq.~(\ref{eq:thematrix}) are given as Eq.~\eqref{eq:matrixev}. Any eigenvalue with a positive real part corresponds to a perturbation that grows with time, i.e., an instability. Examining Eq.~(\ref{eq:matrixev}), we find that there are two conditions under which such instabilities exist, which we detail below.

{\it Positive Real Eigenvalue.} 
A positive real eigenvalue corresponds to a pure growing exponential in the neighbourhood of the equilibrium, so the system will immediately leave the state. As such, we term such solutions `unstable'. They occur when the contents of the outer square root in Eq.~(\ref{eq:matrixev}) are real and greater than 2, which happens when

\begin{equation}
(1+A_{1}B_{1})(1+A_{2}B_{2})<A_{1}A_{2}C^{2},
\end{equation}

\noindent which is equivalent to Eq.~\eqref{eq:unstable}.

{\it Complex Eigenvalue with Positive Real Part.}
If the contents of the inner square root in Eq.~(\ref{eq:matrixev}) are negative, the eigenvalues will be complex, corresponding to perturbations that oscillate, as well as grow or decay exponentially, with time. In this case, their real and imaginary parts, i.e., the exponential growth rate and angular frequency of the perturbations, take the form of Eqs.~\eqref{eq:angfreq} and \eqref{eq:growth}, respectively. We can see that the two eigenvalues for which $R$ takes the + sign will satisfy $R>0$, and so correspond to growing oscillations, when

\begin{equation}
A_{1}A_{2}(B_{1}B_{2}-C^{2})>[2+\frac{1}{2}(A_{1}B_{1}+A_{2}B_{2})]^{2},
\end{equation}

\noindent Applying this to our expression for $\Omega$, we find that, for all growing oscillations,

\begin{equation}
|\Omega|>\sqrt{1+\frac{1}{2}(A_{1}B_{1}+A_{2}B_{2})},
\end{equation}

\noindent with equality when the oscillations are marginal, i.e., when $R=0$.

\bibliography{basename of .bib file}

\end{document}